\documentstyle[prl,tighten,aps,epsf,multicol,amsfonts,times]{revtex}

\newcommand{\thvec}[3]{\left(\begin{array}{c} #1 \\ #2  \\ #3 \end{array}\right)}
\newcommand{\fvec}[4]{\left(\begin{array}{c} #1 \\ #2  \\ #3 \\ #4 \end{array}\right)}
\newcommand{\thmat}[9]{\left(\begin{array}{ccc} #1 & #2 & #3 \\  #4 & #5 & #6 \\ #7 & #8 & #9 \end{array}\right)}
\draft
\begin{document}
\author{S. Virmani$^{1,2}$ and M.B. Plenio$^{2}$}
\title{Construction of extremal local positive operator-valued measures under symmetry}
\address{$^{1}$ Gruppo di Ottica Quantistica, Dip. di Fisica `A. Volta', Universit\`{a} degli Studi di Pavia, via Bassi 6, I-27100 Pavia, Italy}
\address{$^{2}$ QOLS, Department of Physics, Blackett Laboratory, Imperial College, Prince Consort Road, London SW7 2BW, UK}
\date{\today}
\maketitle

\begin{abstract}
We study the local implementation of POVMs when we require only
the faithful reproduction of the statistics of the measurement
outcomes for all initial states. We first demonstrate that any
POVM with separable elements can be implemented by a separable
super-operator, and develop techniques for
calculating the extreme points of POVMs under a certain
class of constraint that includes separability and PPT-ness.
As examples we consider measurements
that are invariant under various symmetry groups (Werner, Isotropic,
Bell-diagonal, Local Orthogonal), and demonstrate
that in these cases separability of the POVM elements is equivalent to
implementability via LOCC. We also calculate the extrema of these classes
of measurement under the groups that we consider, and give explicit LOCC protocols
for attaining them. These protocols are hence optimal methods for locally discriminating
between states of these symmetries. One of many interesting consequences is that
the best way to locally discriminate Bell diagonal mixed states is to perform a 2-outcome POVM
using local von Neumann projections. This is true regardless of the cost function,
the number of states being discriminated, or the prior probabilities. Our
results give the first cases of local mixed state discrimination that
can be analysed quantitatively in full, and may have application to other
problems such as demonstrations of non-locality, experimental entanglement
witnesses, and perhaps even entanglement distillation.
\end{abstract}


\begin{multicols}{2}

% \section{Structure of Paper}
% \begin{itemize}
%    \item Intro
%    \item General Techniques\\
%            a) Sep op $\Leftrightarrow$ Sep POVM's\\
%            b) General tricks for extremal points
%    \item Local state discrimination
%        a) Intro\\
%        b) Isotropic states\\
%        c) Werner states\\
%        d) Bell states\\
%        e) OO states
%    \item Comments and conclusions
% \end{itemize}

\section{Introduction}

The nature of local and global quantum operations plays
a key role in Quantum Information theory \cite{somebasicarticles}. However, although
global quantum operations are well characterised \cite{Kraus},
there is still no convenient mathematical
classification of the so called LOCC transformations (operations
that can be implemented locally with classical communication). In recent years some progress
has been made by trying to understand the non-local properties
of unitaries and by trying to implement
operations with a minimum amount of prior entanglement \cite{Eisert JPP00,Collins LP01,ciracgang}.
Nevertheless, a more general answer remains elusive, and so alternative classes of operations have been defined that can be more
easily characterised. Perhaps the most interesting examples are the Separable and PPT operations introduced by Rains \cite{Rains 99,Rains 01}.
These classes of operations (see section II for definitions) are mathematically
more tractable than the LOCC class, and have been particularly
useful as tools for investigating the entanglement of distillation and local state discrimination \cite{Rains 01,hb1}.
They have also been used to extend the conventional definitions of
entanglement measures \cite{Benndist,entanglementmeasures},
giving indications that quantum state transformations
can adopt a simplified structure under the class of PPT operations \cite{Audenaert PE 02}.

The separable and PPT operations will also play an important
role in this work, where we investigate the local implementation
of quantum measurements. This question is significant as
in many protocols for quantum communication it is important to know how a distantly
separated Alice and Bob may be able to gain information
about quantum states using only local operations
and classical communication. Examples include entanglement
distillation and cryptographic scenarios \cite{hb1,Benndist,Volldist}. The issue of local measurements is also
important for efforts to locally infer some
form of entanglement, such as in developing tests
of non-locality, or the local detection of inseparability \cite{guhne}. However, there is perhaps a much more
fundamental reason for the investigation of local measurements: in all uses of quantum states,
all we ever `see' at the end of the day are measurement outcomes, so one of the most
important things to know is what kinds of statistics
are possible according to the classes of operations that we
are restricted to. As the LOCC operations are an important
paradigm in quantum information, it is important to know
what kinds of measurement outcomes we can obtain with them.

Most of the discussion of local measurements in recent years has
been directed particularly towards local state discrimination
\cite{wal1,wal2,Virmani SPM 01,HSS,ghosh}. However, in this work
we pay more attention to the actual implementation of measurements
locally. We will be inspired by initial steps in this direction
taken by \cite{hb1,hb2}, where both the concepts of PPT-operations
and symmetry were used to design a certain type of cryptographic
scheme for Werner states of dimensions $2^{n} \times 2^n$ (work
that was extended to the multiparty setting in \cite{eggeling}).
Unlike those investigations, however, we will not be directly
interested in any particular application or cost functions, as our
main aim will be to obtain techniques for deciding the local
implementability of measurements, with particular regard to
finding extreme points. Some of the methods are quite general, and
have implications for finding extremal POVMs under constraints
other than local implementability.
%We will first discuss the
%correspondence between the entanglement properties of a given
%POVM's elements and the implementability of the measurement using
%separable or PPT operations \cite{Rains 01}. We then go on to make
%use of symmetry to completely classify the sets of local
%measurements that are invariant under some local unitary groups
%\cite{Vollbrecht W 01}, at the same time deriving the extreme
%points. The results are of prime importance for the local
%discrimination of states with these symmetries. A particularly
%interesting example is given by the Bell symmetries, where we show
%that the best way to locally discriminate Bell diagonal states is
%just to perform a 2-outcome POVM using local von Neumann
%projections. This is true regardless of the cost function, the
%number of states being discriminated, or the prior probabilities.

This paper is structured as follows. In section II we discuss the
connection between the entanglement of POVM elements and
implementability via the different classes of operations
introduced by Rains \cite{Rains 01}. In sections III and IV we use
these ideas to derive the entire (convex) set of local
measurements that satisfy Isotropic and Werner symmetries, also
giving the extreme points. These particular symmetries are
amenable to an elegant form of solution that is not possible for
more complicated symmetry groups, where we will require stronger
methods. So in section V we discuss general techniques for
deriving the extrema of POVMs under a quite general class of
constraint that includes the constraints of separability and
PPT-ness. In section VI we apply these techniques to Bell and $OO$
\cite{Vollbrecht W 01} symmetries, deriving the sets of locally
implementable measurements and their extreme points. The Bell
symmetries are particularly interesting, as we show that the best
way to locally discriminate Bell diagonal states is just to
perform a 2-outcome POVM using local von Neumann projections. This
is true regardless of the cost function, the number of states
being discriminated, or the prior probabilities. Finally in
section VII we summarise and discuss the implications of our
results.

\section{Entanglement classification of POVM elements and relationship to
classes of operations}

We begin by investigating the relationship between the
entanglement properties of POVM elements and the resources
required to implement the POVM. The key result of this section
will be the demonstration that the separability (PPT-ness) of the
POVM elements is equivalent to the implementability of the
corresponding POVM by separable (PPT) operations.
% Applications of
% this result are then discussed in the remainder of this section.

Let us begin by reviewing the concepts of PPT and Separable operations.
The set of PPT operations $P$ is the set of
completely positive (CP) transformations that remain completely
positive when conjugated with the partial transposition operator
$\Gamma$, i.e. both $P$ and $\Gamma \circ P \circ \Gamma$ are
completely positive. It is irrelevant whether the partial
transposition is taken as transposition of Bob's or Alice's
system, as long as the choice is fixed. The set of Separable
operations is the class of operations $S$ that can be written as:
\begin{equation}
    S(\rho) = \sum_{n} A_{n} \otimes  B_{n} \rho A^{\dag}_{n} \otimes B^{\dag}_{n}.
\end{equation}
It is not too difficult to verify \cite{Rains 99} that all LOCC
operations are separable, and all separable operations are PPT. An
example of a CP operation that is not PPT is the creation of an
NPT state from a PPT one \cite{Rains 99}. The PPT and Separable
operations can also be characterised in an interesting way in the
light of the Jamiolkowsky isomorphism \cite{Jam,ciracgang,Rains
99,Rains 01,hb1,hb2}.

This classification of operations has allowed interesting bounds
to be derived in the distillation and creation of entangled states
\cite{Rains 01,Audenaert EJPV 01}, and is also useful for the
study of local quantum state discrimination \cite{hb1,hb2}. In the
following, we will be interested particularly in the question of
when a measurement procedure can be implemented by LOCC. We will
not be concerned with the residual states left over after any
measurement process, so a POVM will be considered to be `locally
implementable' if we can perform a measurement using only local
operations and classical communication that allows exactly the
same statistical inferences to be made. It is important to note
the differences between our use of the word `local' and that of
Beckman et. al. \cite{Preskill} and Groisman et. al.
\cite{Reznik}. Those authors are motivated by the implications of
quantum operations and observations for causality, and hence they
do allow the use of prior entanglement but disallow the use of
classical communication.

In our context the classification introduced by \cite{Rains
99,Rains 01} has a particularly elegant structure, as the PPT and
separable definitions of operations directly map to the
definitions of PPT and separable states \cite{Werner 89,peres}. A
given POVM described by a set of elements $\{M_{k}|k=1..N, M_{k}
\geq 0, \sum M_{k} = \openone \}$ can be implemented by a PPT
operation iff all the elements (after normalisation) correspond to
PPT states, and can be implemented by a separable operation iff
all the elements correspond to separable states:
\begin{eqnarray}
    \mbox{can do POVM by Separable ops} &\Leftrightarrow & \mbox{elements separable} \label{sep} \\
    \mbox{can do POVM by PPT ops} &\Leftrightarrow & \mbox{elements PPT}
\end{eqnarray}
The fact that all measurements implemented by separable operations
must involve separable POVM elements was noted in \cite{hb1}. The
fact that a measurement can be implemented by a PPT operation iff
the POVM elements are PPT was noted by \cite{hb2,prains}. As far
as we are aware, however, the fact that all POVMs with separable
elements can be implemented by a separable operation has not been
noted anywhere despite its relatively simple proof. For
completeness we therefore present the following proof of
correspondence (\ref{sep}):

\smallskip

\noindent {\bf Observation 1.} A given POVM can be implemented by
a separable superoperator iff the POVM elements associated with
all $N$ outcomes $\{M_{k}|k=1..N, M_{k} \geq 0, \sum M_{k} =
\openone \}$ correspond to separable states, i.e. the POVM can  be
implemented by a separable superoperator iff the density matrices
defined by $ M_{k}/ \mbox{tr} \{ M_{k}\}$ are separable.

\smallskip

\noindent {\bf Proof.} Any POVM can be realised as a superoperator where the
subscripts $n$ on the Kraus operators have been recorded
classically. Any separable operation will correspond to POVM elements where the
subscripts $n$ will be grouped into sets $S_{k}$ corresponding to
the different inferences $k$. Hence the expectation values of any
POVM element $M_{k}$ can be written:
\begin{eqnarray}
    \mbox{tr}\{M_{k}\rho\} &=&\sum_{n \in S_{k}}\mbox{tr}\{ A_{n} \otimes  B_{n}
    \rho A^{\dag}_{n} \otimes B^{\dag}_{n}\} \\
    &=& \sum_{n \in S_{k}}\mbox{tr}\{ A^{\dag}_{n} A_{n} \otimes  B^{\dag}_{n} B_{n} \rho
    \}.
\end{eqnarray}
Hence we can write that:
\begin{equation}
    M_{k} =\sum_{n \in S_{k}} A^{\dag}_{n} A_{n} \otimes  B^{\dag}_{n} B_{n}
\end{equation}
As this sum is completely in terms of positive operators, we have
that $ \mbox{tr}\{M_{k}\}=\sum_{n} \mbox{tr} \{ A^{\dag}_{n} A_{n}\}  \mbox{tr} \{
B^{\dag}_{n} B_{n} \}$. This means that we can also write:
\begin{eqnarray}
    \rho_{k} &:=& {M_{k} \over \mbox{tr}\{M_{k}\}} \nonumber\\
    &=&
    \sum_{n \in S_{k}} \frac{ \mbox{tr}\{ A^{\dag}_{n} A_{n} \} \mbox{tr}\{ B^{\dag}_{n} B_{n} \} }{ \mbox{tr}\{M_{k}\} }
    \frac{A^{\dag}_{n} A_{n}}{ \mbox{tr}\{ A^{\dag}_{n} A_{n} \} } \otimes \frac{ B^{\dag}_{n}
    B_{n}}{ \mbox{tr}\{ B^{\dag}_{n} B_{n} \} }. \nonumber
\end{eqnarray}
Hence we see that each POVM element from a separable operation
must correspond to a separable state. To see the converse, simply
work backwards through the above procedure, using the fact that
any positive operator $X$ has a decomposition $X=Q^{\dag}Q$. The
generalisation to multi-party systems is straightforward
$\square$.

\smallskip

Observation 1 is extremely powerful, and can be used to provide
simple derivations of many results. For example we can immediately
see that an entangled pure state cannot be perfectly locally
discriminated from its orthogonal complement, as this would
require a POVM with non-separable elements (see also \cite{HSS}). A
similar argument also shows that we cannot perfectly locally
distinguish an Unextendible Product Basis (UPB) \cite{UPB} from an equal
mixture of the pure states in its orthogonal complement, adding to
the other intriguing discrimination properties of such bases \cite{UPB}.

Is it possible that all separable discrimination protocols can
be implemented locally? Unfortunately this is definitely not the
case, as has been observed from the example of non-locality
without entanglement presented in \cite{nonlocalitywithoutentanglement}
(see also \cite{wal2}). The authors of that
paper present sets of orthogonal separable states that cannot be
discriminated perfectly using only local operations. If
all separable POVMs could be implemented locally then those sets
of orthogonal states could be discriminated perfectly using only
local operations. As this is not the case, this means that
not all separable operations can be implemented locally.

Nevertheless, the constraint that any local POVMs must be
separable is still quite a strong one, and in some simple cases of
high symmetry we will see that it is exactly equivalent to the local
implementability of POVMs. The results thereby allow the
construction of
optimal local discrimination protocols
for any number of states of such symmetries under any cost function for any prior
probabilities. The ideas may also have other applications in
cryptographic schemes, demonstrations of non-locality,
and the local detection of entanglement.
We will begin by considering the {\it Isotropic}
symmetries.

\section{Local observations with Isotropic Symmetries}

% We then go on to consider the larger class of POVMs invariant
% under the {\it Local Orthogonal group} of transformations of the
% form $O \otimes O$. We construct the extreme points of the set of
% PPT POVMs with these symmetries and show how they can be obtained
% by local protocols. This means that any such POVM can be locally
% matched, thereby allowing relatively easy optimisation of local
% discrimination of two $O \otimes O$ symmetric states.

In this section we begin by analysing the so called {\it Isotropic}
symmetries. The class of measurements invariant under this group
is actually a subset of the $OO$ symmetric measurements that we
will consider later in the paper. However, we consider the Isotropic and Werner
cases independently in this section and the next, as they afford
a particularly elegant form of solution and demonstrate the usefulness
of the partial transposition mapping between Isotropic and Werner states.

% Suppose that we have been given one of several quantum states
% $\rho_{i}$, with some arbitrary prior probabilities $w_{i}$, and
% that we wish to determine which state we have been given. Denoting
% the identity operator on the total Hilbert space by $I$, we need
% to perform some POVM $\{M_{k}|k=1..N, M_{k} \geq 0, \sum M_{k} = \openone
% \}$. This POVM will give us outcomes $k$ with probabilities
% \begin{equation}
%    p_{k,i} = w_{i}\mbox{tr}[M_{k}\rho_{i}].
% \end{equation}
% For each outcome $k$ we will make an inference about the original
% state depending upon the structure of the probabilities $p_{k,i}$
% and $w_{i}$, and whatever discrimination figure of merit we choose
% to optimise.

The {\it Isotropic} states are those that commute with all local unitaries
of the form $U \otimes U^{*}$, where the $^{*}$ denotes complex
conjugation in a fixed local basis. Any isotropic state
$\sigma(f)$ on $C^{d} \otimes C^{d}$ can be written as:
\begin{equation}
    \sigma(f)= f|+\rangle\langle +| + (1-f){ \openone - |+\rangle\langle +|  \over d^{2} - 1 },
\end{equation}
where $f \in [0,1]$, and $|+\rangle$ is a canonical maximally
entangled state $|+\rangle = \sum |ii\rangle/\sqrt{d}$. In the local
observation of such states we can use standard symmetry
arguments to restrict our attention to POVMs with elements of
the form:
\begin{eqnarray}
    M_{k} = a_{k}|+\rangle \langle +| + b_{k}(\openone - |+\rangle \langle +|).
\end{eqnarray}
As we require these elements to form a valid POVM, they must satisfy
the constraints:
\begin{eqnarray}
    a_{k} & \geq & 0 \nonumber \\
    b_{k} & \geq & 0 \nonumber \\
    \sum_{k}  a_{k} & = & 1 \nonumber \\
    \sum_{k}  b_{k} & = & 1 . \label{povcon}
\end{eqnarray}
It is well known that isotropic states are separable iff they are
PPT (see e.g. \cite{Vollbrecht W 01}), and so it is easy to
compute the additional constraint that each $M_{k}$ must be
separable from $M_{k}^{\Gamma}\ge 0$:
\begin{equation}
    (d + 1) b_{k} \geq  a_{k} . \label{sepcon}
\end{equation}
We will now see that there is a local protocol that can be used to
attain any Isotropic POVM satisfying equations (\ref{povcon}) and
(\ref{sepcon}). The protocol consists of two steps:
\begin{itemize}
    \item Alice and Bob perform an isotropic twirl \cite{isotwirl}, $T(\rho)$. This step is
superfluous if the states to be observed are isotropic anyway. It is only
included to make the total measurement exactly equal to the POVM satisfying (\ref{povcon}) and (\ref{sepcon}).
    \item Then Alice and Bob perform a measurement with elements $N_{k}$ described by:
\begin{equation}
    N_{k} = \sum_{i=1}^{d} |i\rangle \langle i| \otimes [x_{k} |i\rangle \langle i|
    + y_{k}(\openone _B - |i\rangle \langle i|)] \label{isomethod}
\end{equation}
with
\begin{eqnarray} x_{k} &=& a_{k} \nonumber \\ y_k &=&{(d+1)b_{k} - a_{k}\over d }, \end{eqnarray}
\end{itemize}
where $|i\rangle$ is a pure state from the computational bases of
Alice $\&$ Bob, and $\openone _B$ refers to the identity on Bob's
space.

The probability that Alice and Bob will find outcome $k$ from
the above procedure given an input state $\rho$ is given by:
\begin{equation}
    \mbox{tr}\{ N_{k} T(\rho) \} = \mbox{tr}\{ T(N_{k}) \rho \}.
\end{equation}
It is not difficult to verify that $M_k =$T($N_k$) for all $k$,
and so our two step protocol gives exactly the same statistics for
each $k$ as the original measurement. Moreover, examination of the
form of the $N_k$ shows that they correspond to a local
measurement involving only one way communication, as long as the
condition (\ref{sepcon}) holds (otherwise the $N_{k}$ will not be
positive). Therefore any POVM consisting of PPT Isotropic elements
can be attained locally. The transformation above also gives a
direct derivation of the extreme points of this class of POVMs.
The mapping from the PPT ($a_{k},b_{k}$) to the ($x_{k},y_{k}$) is
linear and invertible. The $x_k$ and $y_k$ form two independent
probability distributions, and so the extreme points will just
correspond to situations when the $x_{k}, y_k$ take on values of
$\{0,1\}$. This observation allows relatively straightforward
optimisation of local discrimination of isotropic states under
essentially all cost functions.

It is also worthwhile noting that there are many other possible
local protocols that can match the PPT isotropic POVMs. This is
because the above protocol essentially makes full use of the
maximally entangled component of the isotropic states to leave Bob
in a residual mixture of some pure state with the identity, which
can then be used to discriminate between the original isotropic
states. Indeed, any measurement that Alice can perform on a
maximally entangled state that would leave Bob with a known pure
state can be used to construct a similar protocol. One example is
teleportation. Suppose Alice and Bob first teleport a known pure
state $|\psi\rangle$ using whichever of the isotropic states that
they share. Then if Bob performs a POVM defined by elements $x_{k}
| \psi \rangle \langle \psi| + y_{k}(\openone _B - |\psi \rangle
\langle \psi|)$ then they will also achieve the same POVMs as the
$N_k$ defined above.

\section{Local observations with Werner Symmetries}

Having calculated the set of locally implementable isotropic
POVMs, let us see why we can almost immediately write down the set
of locally implementable Werner POVMs \cite{Werner 89}. The {\it Werner}
states are those that commute with all local unitaries
of the form $U \otimes U$. Any Werner symmetric POVM can be written
with elements of the form:
\begin{eqnarray}
    M_{k} = a_{k} P_A + b_{k} P_S,
\end{eqnarray}
where $P_A$ and $P_S$ are the so called {\it antisymmetric} and
{\it symmetric} projectors respectively \cite{Werner 89}.

The argument here follows from the partial transposition mapping
between Isotropic and Werner operators that was discussed in
\cite{Vollbrecht W 01}. The authors of that paper point out that
if a certain class of operators is invariant under a given group
of local unitary operations $U(\gamma) \otimes V(\gamma)$, then
the partial transpositions of those operators will be invariant
under the `partially conjugated' group $U(\gamma) \otimes
V^{*}(\gamma)$ (modulo a problem of phases, see \cite{Vollbrecht W 01}). This
means that if we have a set of PPT POVM elements that is invariant
under one local symmetry group,
\begin{equation}
    \{M_{k}| \sum M_k = \openone, M_k \geq 0, M_k^{\Gamma} \geq 0\}
\end{equation}
then the partial transposition of these elements will form
another POVM that is a PPT POVM invariant under the
partially conjugated group, and vice versa. Indeed the
extremal PPT measurements under one group of local unitaries are in one-to-one
correspondence with extremal PPT measurements under the `partially
conjugated' group:

\smallskip

\noindent {\bf Theorem 2.} If $\{M_k\}$ forms an extremal PPT POVM for
one local symmetry group, then the partial transposition $\{M_k^{\Gamma}\}$
forms an extremal PPT POVM for
the partially conjugated symmetry group.

\smallskip

\noindent {\bf Proof.} If a POVM $\{M_k | k=1...n\}$ is PPT and
{\it not} extremal, then it is possible to write:
\begin{equation}
\{M_k | k=1...n \} = \{p M_k^{1} + (1-p) M_k^{2} | k=1...n \}
\end{equation}
for some probability $p$, where $\{M_k^{1} | k=1...n\}$ and
 $\{M_k^{2} | k=1...n\}$ are also PPT POVMs. Therefore the partial
transposition of this equation is also true:
\begin{equation}
\{(M_k)^{\Gamma} | k=1...n \} = \{p (M_k^{1})^{\Gamma} + (1-p) (M_k^{2})^{\Gamma} | k=1...n \},
\end{equation}
where $\{(M_k^{1})^{\Gamma} | k=1...n\}$ and
 $\{(M_k^{2})^{\Gamma} | k=1...n\}$ are also PPT POVMs.
So the partial tranposition of a PPT POVM is
extremal iff the PPT POVM itself is extremal
$\square$.

\smallskip

This convenient partial transposition connection is precisely the relationship between
PPT POVMs of Werner symmetry and PPT POVMs of Isotropic
symmetries.  Hence all implications
discussed above for the local discrimination of isotropic states
also immediately apply to local discrimination of Werner states.
In particular, the partial transposition of the local
protocols given for Isotropic symmetries above will allow us to
obtain any PPT POVM with Werner symmetries. To be explicit,
let us denote a Werner-symmetric POVM by elements:
\begin{equation}
   M_{k} = a_{k} P_A + b_{k} P_S .
\end{equation}
The condition that the POVM be PPT forces us to impose the requirement
that:
\begin{equation}
b_k \left( { d+1 \over d-1}\right) \geq a_k
\end{equation}
in addition to the other positivity and completeness constraints. Any such POVM
can be attained by the partial transposition of a PPT isotropic measurement,
and so can be attained by the `partial transposition' of the local measurement
used to attain that particular isotropic POVM, i.e.:
\begin{itemize}
    \item First Alice and Bob perform a Werner twirl \cite{isotwirl}, $T(\rho)$. Again, this step is
superfluous if the states to be observed are Werner anyway.
    \item Then Alice and Bob perform a measurement with elements $N_{k}$ described by:
\begin{equation}
    N_{k} = \sum_{i=1}^{d} |i\rangle \langle i| \otimes [x_{k} |i\rangle \langle i|
    + y_{k}(\openone _B - |i\rangle \langle i|)]
\end{equation}
with
\begin{eqnarray} x_{k} &=& { (1-d) a_k + (d+1) b_k \over 2} \nonumber \\ y_k &=& a_k . \end{eqnarray}
\end{itemize}
It can readily be verified that the protocol matches any PPT Werner POVM.

In the other symmetry groups that we will consider in this paper,
no elegant and simple solution along the lines of the one found
for the Isotropic/ Werner states can work. The reasons for this
will be discussed in more detail in the section on $OO$ symmetries
and in the appendix. Consequently, for other symmetry groups we
have to adopt a different approach. Here we choose to construct
the extremal points of the (convex) set of PPT POVMs for the
groups that we consider, and then try to find local protocols that
match these extrema. We then need to invoke a theorem of convex analysis
which states that any convex compact set is the (closed) convex hull of its extreme points \cite{KM},
and hence if we can show that the extremal points
of the PPT measurements can be obtained locally, then by convexity so
can the whole set (for a finite number of outcomes
the set of PPT POVMS is clearly compact). In the cases that we consider here our search
for local protocols will always be successful. In fact, the most
difficult exercise is usually the construction of the extremal
points. The task is made tractable by some general techniques that
we will discuss next.

\section{General Techniques and Notation}

In this section we will explain the major tools that
we will use throughout the rest of the paper. Sometimes we will
describe these techniques in quite general
language, as there may be applications
to other problems.

The first technique, that we will also use later
in this subsection, is a quite common
method for determining constraints on the extrema of
convex sets. Suppose that we have a convex set $A$, and
we are trying to decide whether a candidate element $s$
of the set is extremal. When we write
the phrase `by the following perturbation', this will mean that we
consider the following common approach to deciding extremality.
We can consider perturbations of $s$ by a small
amount $\delta$. We then try to decompose
the candidate extremum $s$ in the following way:
\begin{eqnarray}
    s &=& {1 \over 2} s(+) + {1 \over 2} s(-) \nonumber \\
    s(+) &=& s + \delta \nonumber \\
    s(-) &=& s - \delta .
\end{eqnarray}
In order that $s$ be extremal, we will require that at least one
of $s(+)$ and $s(-)$ be outside the set $A$. In our specific
application, we will usually choose $\delta$ such that the
separability and completeness of the POVM is maintained,
and so we will have to infer that either $s(+)$ or $s(-)$
be non-positive, and this will allow us to derive
strong constriants of the form of the extrema. So when we write `by the perturbation $\delta$ we
can see that the extremal points must be of the form...' we are in
fact referring to the inferences that can be made by the
above process.

Now let us consider how we can use some of the structure of sets of
separable POVMs in order to simplify the hunt for extremal points.
The most important feature that we will rely on is the fact
that the properties of positivity, separability and PPT-ness are closed under
linear combinations with non-negative scalars. Some of the methods that we will employ
can be used to extremise POVMs for any other constraints that have this property,
and not only PPT-ness or separability.
In light of this we will find it convenient to make the
following general definition:

\smallskip

\noindent {\bf Definition 3.} A constraint on two operators $M_1 , M_2$ will be called
{\it homogeneous} if whenever $M_1$ and $M_2$ satisfy the constraint then
so does $\alpha M_1 + \beta M_2$, for all non-negative scalars $\alpha, \beta$.
Hence both Separability and PPT-ness are homogeneous constraints. We will
sometimes use the symbol {\bf H} to denote {\it any} set of homogeneous constraints
under consideration, not necessarily only PPT-ness or separability.

\smallskip

\noindent {\bf Definition 4.} A POVM $\{M_k | k=1...n, M_k \geq 0,
\sum M_k = \openone \}$ satisfying any extra {\it homogeneous}
contraints {\bf H} under consideration will be called a {\it
feasible} measurement.

\smallskip

The set of feasible POVMs is clearly a convex set itself, in that if $\{ M_{k}^{1} | k=1....n \}$
and $\{ M_{k}^{2} | k=1....n \}$ are feasible POVMs, then
so is $\{ p M_{k}^{1} + (1-p) M_{k}^{2} | k=1....n \}$ where $p \in [0,1]$.
Hence we can ask questions about what the extremal POVMs are, and how to find them.
One particularly important result in this direction is the following:

\smallskip

\noindent {\bf Theorem 5.} In any extremal feasible POVM, the non-zero operator elements must be linearly
independent.

\smallskip

\noindent {\bf Proof.} The proof can be
adapted straightforwardly from a proof of this statement for global POVMs
 by Lo Presti $\&$ D'Ariano \cite{lopresti}, and indeed their
proof is essentially true for {\it any} constraints that are
closed under multiplication by non-negative
scalars $\square$.

\smallskip

This theorem will be particularly useful as it will allow us to
restrict the number of non-zero outcomes that we need to consider.
It also provides another reason why in the case of Isotropic/Werner
symmetries the extremal PPT POVMs have essentially two non-zero
outcomes, simply because under these symmetries there are at most two
non-zero linearly independent POVM elements.

The consideration of homogeneous constraints  \cite{oliver} will allow us to
use a technique that involves what we call `basic
vectors'. This should not be confused with the term `basic
solution' that is often used in linear programming problems. In
the rest of this section we will explain exactly what we mean by a
`basic vector'. Let us suppose that we can find a class of elements $\{V(\alpha)\}$
where $\alpha$ represents one or more parameters (discrete or
continuous) with the following properties:
\begin{itemize}
\item Each element from $\{V(\alpha)\}$ is non-negative.
\item Each element from $\{V(\alpha)\}$ satisfies the homogeneous constraints {\bf H} under consideration.
\item All POVM elements $M_k$ satisfy the homogeneous constraints under consideration iff they can
be written in the following way:
\begin{equation}
M_k = \sum_{i} p_i^k V(\alpha_i)
\end{equation}
where the $p_i^k $ are non-negative real numbers.
\end{itemize}
Then the set $\{V(\alpha)\}$ will be referred to as a set of {\it
basic vectors}. The coefficient $p_i^k$ will be referred to as the {\it
weight} of the basic vector $V(\alpha_i)$ in $M_k$. Of course this
decomposition, and hence the weights, need not be unique. Such
sets of basic vectors can be very useful tools for deriving the
set of extreme points, and in subsequent sections we will prove a
few lemmas that will demonstrate their power. The source of this
power is the fact that in the `basic vector' approach, each feasible POVM
element is automatically decomposed into a sum (with non-negative weights) of operators
that automatically satisfy the homogeneous constraints that we are
considering. So we are free to try to perturb any non-zero
weights by small amounts without violation of our constraints.
This allows us to derive strong limitations on
which weights can be non-zero in an extremal measurement. In particular
the following two Lemmas prove to be useful:

\smallskip

\noindent {\bf Lemma 6.} In an extremal feasible POVM, no basic vector
can have non-zero weight in more than one element $M_k$ of the POVM.

\smallskip

\noindent {\bf Proof.} This can be proved along similar lines to the proof
of the following lemma $\square$.

\smallskip

\noindent {\bf Lemma 7.} If a feasible POVM is extremal,
then any set of linearly dependent basic vectors must
have non-zero weights in at most one element $M_k$.

\smallskip

\noindent {\bf Proof.} Suppose that we have a set of basic vectors $\{
{V}_{m} |m=1...L \}$ such that:
\begin{equation} \sum_{m} \lambda_{m}
{V}_{m} = 0  \end{equation}
for some set of non-zero $\lambda_{m}$'s.
We can take the Hermitian conjugate of this equation and average it
with the original  to give:
\begin{equation} \sum_{m} {1 \over 2} \lambda_{m}
{V}_{m} +  {1 \over 2} \lambda_{m}^{*}
{V}_{m} = \sum_{m} \mbox{Re}\{\lambda_{m}\}
{V}_{m} = 0 .\end{equation}
Hence the $\lambda_{m}$s can be taken to be real
without loss of generality (this is unless all the original
$\lambda_m$ are purely imaginary, but in that case we can
just divide the equation through by $\sqrt{-1}$). Suppose that the basic vector ${V}_L$ has non-zero
weight in element $M_2$, whereas the other basic vectors of the
set have non-zero weight in $M_1$. Then we have
that:
\begin{equation} \sum_{m \neq L} {\lambda_{m} \over \lambda_{L}} {V}_{m}
 = - {V}_{L}  \end{equation}
We define two new POVMs
$s(\pm )$ with exactly the same elements and basic weights as the
original POVM except for the following changes:
\begin{equation}
    p_{m \neq L}^{1}(\pm) = p_{m}^{1} \pm \delta {\lambda_m \over \lambda_L}~~~;~~~ p_{L }^{2}(\pm) = p_{L }^{2} \pm \delta
\end{equation}
The two new POVMs average to give the original, are complete
by construction, and if the
$\delta$ is chosen small enough, will not violate any of the
homogeneous constraints or positivity. The same
argument can easily be extended to the case where more than one
of the vectors from the set is placed in $M_2$. Therefore we
require that any set of linearly dependent basic vectors contains
non-zero weights in at most one fixed element of an extremal POVM  $\square$.

\smallskip

The exact utility of the basic vectors approach will become
clearer in the examples that we tackle later, where we will
explicitly construct sets of basic vectors. Unfortunately
it is not always easy to do this construction. Nevertheless, there is one approach based
upon construction of two outcome extremal POVMs that
can allow us to draw some interesting general conclusions.  Let us consider the set of feasible two outcome POVMs:
\begin{equation}
\{ M, \openone -M \}.
\end{equation}
Each such measurement is characterised by just one element $M$, as
the other is fixed by completeness. Suppose then that we can
somehow find the extreme points of this set, and they are given by
the set $\{M(\alpha )\}$, where $\alpha $ represents some
parameters (discrete or continuous) that label the extrema. Then
it is clear that $\{M(\alpha )\}$ will form a set of basic vectors
for feasible POVMs with {\it any} number of outcomes. This is
because any particular element $M_i$ of an $N$-outcome POVM can be
viewed as a member of a two outcome POVM $M_i , \openone -M_i $,
where the remaining elements have been grouped together as one
outcome, and so any element of a feasible measurement can always
be written as a convex combination of the $\{M(\alpha )\}$. This
feature will be particularly
 useful to us in the construction of the separable $OO$ symmetric
extrema, as in that case we will be able to construct the
two outcome extrema more easily.

There are also more general conclusions that can be made from this
observation. Suppose that each two-outcome feasible POVM element can be constrained to live
on a $K$-dimensional vector space. If the elements are $m \times m$
Hermitian operators, then there is a natural upper bound to $K$
of $m^{2}$, although this can be reduced in cases of symmetry.
Then by Lemma 7 above, we know that each linearly dependent
set of basic vectors can contribute non-zero weights to at most
one POVM element. As any $K+1$ elements drawn from $\{M(\alpha )\}$
are linearly dependent, we must conclude that in any $K$ non-zero
outcome extremal feasible measurement, each element must be
proportional to one and only one distinct member of $\{M(\alpha )\}$.
Hence we have that:

\smallskip

\noindent {\bf Lemma 8.} In any feasible extremal measurement of $K$ non-zero outcomes,
each POVM element must be proportional to an element of a 2-outcome extremal
feasible POVM.

\smallskip

This result is interesting as it shows how information from 2-outcome optimisations
may be used to draw conclusions about problems with higher numbers of outcomes.
Note that the possible constants of proportionality will be uniquely fixed
by completeness, as each of the $K$ 2-outcome extrema contributing to the
POVM are linearly independent. It is important to note that
there cannot be any extremal feasible measurements with more than $K$ non-zero outcomes anyway
by the linear independence requirement.

We are now in a position to apply these tools to the cases of symmetry that
we will consider in the rest of the paper.

\section{Bell and $OO$ symmetric measurements}

In the previous section we have discussed a number of useful
results that we can use calculate the extreme points of separable POVMs.
Here we will now use these techniques and notations to
derive the set of local measurements of Bell and $OO$ symmetries,
and to characterise the extreme points.

The symmetry groups considered here are all taken from
the examples discussed in \cite{Vollbrecht W 01}. They have
the convenient property that the invariant POVMs have
elements that can be written as a decomposition into a finite
number $n$ of orthogonal projectors $A,B,C...$:
\begin{equation}
    M_i = a_i A + b_i B + c_i C ......
\end{equation}
(i.e. the {\it commutant} \cite{commutant} of the group is abelian). Therefore each
POVM element can be represented by a vector of coefficients $(a_i
, b_i , c_i ,...)^{T}$. As at most $n$ vectors with $n$ components
can be linearly independent, this means that we will be able to
write the most general extremal PPT POVM for the symmetry group
under consideration as an $n \times n$ matrix, where each column
$j$ contains the column vector $(a_j , b_j , c_j ,...)^{T}$
corresponding to the POVM element $M_j$.

Another key feature of the symmetries that we consider is that
checking the positivity of the partial transposition of a
particular POVM element is relatively simple, as the partial
transposition of each invariant POVM element can always be written
in the following way:
\begin{equation}
    M_i^{\Gamma} = a_i' A'  + b_i' B' + c_i' C' ......
\end{equation}
where the $(a_i' , b_i' , c_i' ....)$ form a vector of real
coefficients and the $(A',B',C' ...)$ are a set of mutually
orthogonal projectors. Note that this is not always the case, in
that there can be local symmetry groups with an abelian commutant
whose partial tranposition is no longer abelian. Nevertheless,
there are many interesting symmetries that have this property,
including those of Bell diagonal states that we consider next.

\subsection{Local observation with Bell symmetries}

In this section we will derive the extreme points of the set of
Bell Diagonal PPT POVMs, and then see that they can be
attained locally. We are forced to adopt this route, as no direct and simple solution
in the manner of the Isotropic/ Werner case is possible for the Bell
group. A more detailed explanation for this will be given in the appendix.

First we will explain our notation. The Bell basis is defined as
the orthonormal set of two qubit pure states:
\begin{eqnarray}
    |\Phi^{\pm}\rangle &=& {1 \over \sqrt{2}} (|00\rangle \pm |11\rangle )\nonumber \\
    |\Psi^{\pm}\rangle &=& {1 \over \sqrt{2}} (|01\rangle \pm |10\rangle ),
\end{eqnarray}
and it constitutes a set of projectors that is
invariant under the symmetry group
$\{\sigma_x\otimes\sigma_x,\sigma_y\otimes\sigma_y,\sigma_z\otimes\sigma_z,\openone\otimes\openone\}$
discussed in \cite{Vollbrecht W 01}. Consequently, any POVM element invariant
under the same group can be written as:
\begin{eqnarray}
    M_i &=& a_i |\Psi^{+}\rangle \langle\Psi^{+}| +
    b_i |\Psi^{-}\rangle \langle\Psi^{-}| \nonumber \\
&+&  c_i |\Phi^{+}\rangle \langle\Phi^{+}|
    + d_i |\Phi^{-}\rangle \langle\Phi^{-}|. \label{definemi}
\end{eqnarray}
As we require linear independence between the non-zero POVM
elements, we can restrict our attention to candidate extremal
measurements represented by a $4 \times 4$ matrix, where each
column $i, i=1...4$ has elements given by $\{a_i , b_i , c_i , d_i
\} ^{T}$, the quadruple representing the POVM element $M_i$:
\begin{equation}
    s = \left(\begin{array}{cccc}
    a_{1} & a_{2} & a_{3} & a_{4} \\
    b_{1} & b_{2} & b_{3} & b_{4} \\
    c_{1} & c_{2} & c_{3} & c_{4} \\
    d_{1} & d_{2} & d_{3} & d_{4} \end{array}\right).
\end{equation}
In each column all elements must be non-negative, and each row
must sum to 1 for completeness.

\smallskip

\noindent {\bf Definition 9.} A largest element in a given column will be
referred to as the a {\it maximal} element of the column.
In many cases the maximal element of a column
 will not be unique, in which case we are free to choose
among the possibilities.

\smallskip

The condition that each column
corresponds to a PPT measurement is that a maximal element of a column
must be less than or equal to the sum of the other three
elements of the same column.

\smallskip

\noindent {\bf Definition 10.} We will refer to a column
as being {\it tight} if it has a maximal element {\it equal} to
the sum of the remaining matrix elements in its column.

\smallskip

\noindent {\bf Lemma 11.} In an extremal Bell POVM with a finite
integer $N$ of non-zero outcomes, at least $N-1$ columns must be
tight.

\smallskip

\noindent {\bf Proof.} Suppose that we consider two columns,
say without loss of generality the first two. Suppose that there are two
rows such that both columns contain non-zero
matrix elements. Suppose w.l.o.g. these two
rows are the top two rows. Then we can perturb the POVM
as follows with a small positive $\delta$:

\begin{equation}
    s(\pm) = s \pm \left( \begin{array}{cccc} +\delta & -\delta & 0 & 0 \\
    +\delta & -\delta & 0 & 0 \\ 0 & 0 & 0 & 0 \\  0 & 0 & 0 & 0 \end{array} \right).
\end{equation}

If $\delta$ is small enough, this perturbation maintains
positivity and completeness. Therefore PPT-ness must be violated.
Computing the partial transpose of $M_i$ in eq. \ref{definemi} the
positivity of $M_i^{\Gamma}$ is equivalent to
\begin{eqnarray}
    a_i+b_i &\ge & |c_i-d_i| \label{cond1}\\
    c_i+d_i &\ge & |a_i-b_i| \label{cond2}.
\end{eqnarray}
If PPT-ness should be violated by arbitrarily small
perturbations $\delta$ this implies that at least one of these
conditions has to be an equality. Together with the
positivity of the $a_i,b_i,c_i$ and $d_i$, this implies the tightness of at
least one column.

Alternatively, it could be the case that there are no two rows in
which both columns contain non-zero matrix elements. This can only
be the case if one of the columns contains only two non-zero
matrix elements. In that case both non-zero matrix elements of the
column in question must be maximal in order for it to be PPT, and
so that column must be tight.

We can take any two columns in this argument, and so at least
$N-1$ of the non-zero columns must be tight $\square$.

\smallskip

\noindent {\bf Lemma 12.} Any tight column $j$ can itself be written
in the following way:

\begin{equation}
\sum_{i} p^{j}_{i} P_{i} \fvec{1}{1}{0}{0} ~~ + ~~ q^{j} \fvec{0}{0}{0}{0},
\end{equation}
where the summation runs over all permutation matrices $P_i$ and
the coefficients $\{ p^{j}_{i}, q^{j} \}$ form a probability distribution.

\smallskip

\noindent {\bf Proof.} Let the sum of the matrix elements in the column be
$2\sigma_j$. Then as the column is tight, we must have that its
maximal element is equal to $\sigma_j$. Therefore the column is
majorised \cite{Bhatia} by the column vector:
\begin{equation}
    \fvec{\sigma_j}{\sigma_j}{0}{0}.
\end{equation}
It is a well known result that if a column vector $\vec{g}$ majorises a column
vector $\vec{h}$ then $\vec{h}$ can be written as a convex combination
of permutations of $\vec{g}$ \cite{Bhatia}. Hence we can write that:
\begin{equation}
    \sigma_j \sum_{i} x^{j}_{i} P_{i} \fvec{1}{1}{0}{0} ~~+~~ (1-\sigma_{j})\fvec{0}{0}{0}{0},
\end{equation}
where the $\{ x^{j}_{i}\}$ form a probability distribution. As
no maximal element can be greater than 1 (from completeness),
the $\sigma_j, 1-\sigma_j$ also form a probability distribution,
and so by setting  $ p^{j}_{i} = \sigma_j x^{j}_{i}$ and
 $q^{j}=\sigma_j$ we have a decomposition of the column as stated in the lemma.
Note that the decomposition need not be unique.

\smallskip

\noindent {\bf Lemma 13.} All columns $j$ in a Bell PPT POVM can be expressed as

\begin{equation}
    \sum_{i} p^{j}_{i} P_{i} \fvec{1}{1}{0}{0} ~~ + ~~ q^{j} \fvec{0}{0}{0}{0} + ~~ r^{j} \fvec{1}{1}{1}{1}
\end{equation}
where the $\{p^{j}_{i}, q^{j},  r^{j}\}_{i}$ form a probability distribution.

\smallskip

\noindent {\bf Proof.} Let us for now just consider extremal measurements
of two non-zero outcomes. One of the two non-zero columns of any such measurement must be tight,
let this w.l.o.g. be column 1. Then from completeness and Lemma 12 we can expand column 2 as:
\begin{eqnarray}
 \fvec{a_2}{b_2}{c_2}{d_2} &=&  \fvec{1}{1}{1}{1} -
\left[   \sum_{i} p^{1}_{i} P_{i} \fvec{1}{1}{0}{0} ~~ + ~~ q^{1} \fvec{0}{0}{0}{0} \right] \nonumber \\
 &=&
 \sum_{i} p^{1}_{i} P_{i} \fvec{0}{0}{1}{1} ~~ + ~~  q^{1} \fvec{1}{1}{1}{1} \nonumber \\
 &:=&
 \sum_{i} p^{2}_{i} P_{i} \fvec{0}{0}{1}{1} ~~ + ~~  r^{2} \fvec{1}{1}{1}{1}, \nonumber
\end{eqnarray}
where as before $\{ p^{2}_{i}, r^{2} \}$ also forms a probability distribution.
Hence the theorem holds for all 2 non-zero outcome extremal PPT Bell measurements,
and by convexity holds for the non-extremal ones as well. But then any column from
an $N$ outcome PPT measurement can be viewed as a member of a two outcome
measurement where the remaining columns have been summed. Therefore the
result is true for any column in any Bell PPT POVM.
 $\square$.

\smallskip

A consequence of this lemma is that the set of column vectors of the form:
\begin{equation}
 P_{i} \fvec{1}{1}{0}{0} ~~\mbox{or}~~ \fvec{1}{1}{1}{1}
\end{equation}
forms a set of basic vectors for the Bell Diagonal PPT POVMs. For
a column $j$, the probability $p^{j}_{i}$ or $r^{j}$ will hence be
taken as the weights of the correponding basic vector. We
will refer to as $(1,1,1,1)^{T}$ as the {\it identity basic
vector} as it corresponds to the identity POVM.

\smallskip

\noindent {\bf Lemma 14.} There are no 4 non-zero outcome extremal Bell POVMs.

\smallskip

\noindent {\bf Proof.} In a 4 non-zero outcome extremal POVM, each non-zero
column must have weight from only one basic vector distinct from the
basic vectors contributing to the other columns. Otherwise the linear
independence condition in Lemma 7 will be violated (as
any five 4-component column vectors must be linearly dependent).

If one column contains weight from the identity, then the POVM
cannot be extremal, as it can be expressed as the convex sum
of the identity and a renormalised POVM that consists of
only the remainder from the original.

For four outcomes where no column contains weight
from the identity, a little thought shows that
the only possibility that can maintain completeness
is a POVM given by a row and/or column
permutations of the following:
% \begin{equation}
% \left( \begin{array}{cccc} {1\over 2} & 0 & 0 & {1\over 2} \\
%  {1\over 2} & {1\over 2} & 0 & 0 \\ 0 &  {1\over 2} &  {1\over 2} & 0 \\  0 & 0 &  {1\over 2} &  {1\over% 2} \end{array} \right).
% \label{foor}
% \end{equation}
\begin{equation}
 \left( \begin{array}{cccc} {w} & 0 & 0 & {z} \\
  {w} & {x} & 0 & 0 \\ 0 &  {x} &  {y} & 0 \\  0 & 0 &  {y} &  {z} \end{array} \right),
\label{foor}
\end{equation}
% However, this is not extremal unless one of these columns is zero, as can be verified by perturbing in
% the following way:
% \begin{equation}
% \pm \delta\left( \begin{array}{cccc} +1 & 0 & 0 & -1 \\
%  +1 & -1 & 0 & 0 \\ 0 & -1 &  +1 & 0 \\  0 & 0 & +1 & -1 \end{array} \right).
% \end{equation}
% This perturbation maintains completeness, PPT-ness and positivity if $\delta$ is small
% enough, but as the original POVM can be expressed as an average of the
% two perturbed ones, this means that the original POVM is not extremal.
However, this measurement cannot be extremal as the columns are linearly dependent
To see this, note that we can obtain the zero vector by (a) multiplying each column by a non-zero factor such that each non-zero matrix element has the same value, and then (b) subtracting the 2nd and 4th columns from the sum of the 1st and 3rd.
Therefore, there are no four non-zero outcome extremal PPT Bell POVMs.

\smallskip

\noindent {\bf Lemma 15.} There are no 3 non-zero outcome extremal Bell POVMs.

\smallskip

\noindent {\bf Proof.} As with the proof for four non-zero outcomes, there can
be no column with weight from the identity. Therefore we require
that each column can be decomposed into convex sums of basic
vectors other than the identity, such that no more than four basic
vectors are involved in total. Therefore two columns must be directly
proportional to a basic vector, as each basic
vector can have non-zero weight in at most one column.
Let these two columns be 1 and 2 w.l.o.g.. There are essentially
only two possibilities upto row and column permutations - the
first two columns have a common row in which they both have
non-zero matrix elements, or they have no such common row. The third column
is fixed by completeness. Therefore up to row and column permutations
we have:
\begin{equation}
 \left( \begin{array}{cccc}  x & 0 & 1-x & 0 \\
  x & y & 1-x-y & 0 \\ 0 & y &  1-y & 0 \\  0 & 0 &  1 & 0 \end{array} \right) ~\mbox{or}~
 \left( \begin{array}{cccc} x & 0 & 1-x & 0\\
  x & 0 & 1-x  & 0 \\ 0 & y &  1-y & 0 \\  0 & y &  1-y & 0 \end{array} \right),
\label{3pos}
\end{equation}
but these POVMs cannot be extremal unless they have one further zero
column. For the first matrix we can see this from the following. Consider
the perturbation
\begin{equation}
 \pm \delta \left( \begin{array}{cccc}  +1 & 0 & -1 & 0 \\
  +1 & -1 & 0 & 0 \\ 0 & -1 & +1 & 0 \\  0 & 0 & 0 & 0 \end{array} \right).
\end{equation}
Unless either $x$ or $y$ are zero, this perturbation maintains
completeness, positivity and PPT-ness. Hence the first matrix cannot
be extremal. For the second matrix we can consider the following perturbation:
\begin{equation}
 \pm \delta \left( \begin{array}{cccc}  +1 & 0 & -1 & 0 \\
  +1 & 0 & -1 & 0 \\ 0 & -1 & +1 & 0 \\  0 & -1 & +1 & 0 \end{array} \right).
\end{equation}
This perturbation implies that to be extremal either $x$ or $y$ are zero, or that
the third column is tight. We cannot have three non-zero columns
if $x$ or $y$ are zero, so let us deal with the possibility that the third column
is tight. The third column can only be
tight if one of $x$ or $y$ is 1. Suppose w.l.o.g. that $x=0$, then the second and
third columns become proportional, and so it cannot be extremal. Hence
neither of the possibilities in (\ref{3pos}) can be extremal $\square$.

\smallskip

Therefore there the only extremal POVMs of Bell diagonal form
are those of at most two non-zero outcomes.

\smallskip

\noindent {\bf Theorem 16.} The extremal points of the PPT Bell diagonal
POVMs can be obtained by local protocols. The extremal
points are given by the possible row and column permutations
of the following POVMs:
\begin{equation}
 \left( \begin{array}{cccc} 1 & 0 & 0 & 0 \\
  1 & 0 & 0 & 0 \\ 1 & 0 & 0 & 0 \\  1 & 0 & 0 & 0 \end{array} \right) ~~;~~
 \left( \begin{array}{cccc} 1 & 0 & 0 & 0 \\
  1 & 0 & 0 & 0 \\ 0 & 1 & 0 & 0 \\  0 & 1 & 0 & 0 \end{array} \right).
\end{equation}

\smallskip

\noindent {\bf Proof.} The fact that these are the only extremal points
can be proven using similar arguments to the ones above. They are also locally
attainable, as the the first matrix corresponds to the `do nothing'
measurement, and the other POVMs correspond to POVMs of the form:
\begin{equation}
M = \sigma_{k} \otimes \sigma_{l} [|00\rangle \langle 00| + |11
\rangle \langle 11|] \sigma^{\dag}_{k} \otimes
\sigma^{\dag}_{l}~;~\openone -M
\end{equation}
where the $\sigma_i$ are drawn from the four Pauli matrices. These
POVMs are trivially locally attainable. They simply correspond to
measurements in a product basis, and can be achieved using one way
communication $\square$.

Despite some prior results showing that
the four orthogonal Bell states cannot be perfectly discriminated locally
\cite{ghosh,wal2}, this result is interesting as it shows that there can be a huge
difference between locally and globally obtainable information
 under {\it all} figures of merit, even when we are only dealing
with two qubits. Indeed, as the extremal local measurements involve only
2 non-zero outcomes, there can be at most 1 classical bit of information
obtained by local measurements of Bell diagonal states. However, in the global
case there can be 4 orthogonal Bell diagonal states that can be perfectly
discriminated, thereby allowing a full 2 classical bits of information to be obtained
globally. It might be tempting to propose a bit hiding scheme
on the basis of this observation. However, the scheme would be
susceptible to cheating with a prior shared singlet state (as then
Alice could teleport her particle to Bob), and hence the technology
required to break the scheme is equivalent to that required to
implement it.

\subsection{Local Observations with $OO$ Symmetries}

In the notation of \cite{Vollbrecht W 01}, the OO group is the set
of unitaries of the form $O \otimes O$, where $O$ is an orthogonal
transformation (i.e. $OO^{T}=I_{A/B}$) in some fixed local bases.
Any POVM element invariant under this group can be written in the
form:
\begin{eqnarray}
    a\left[|+\rangle \langle +|\right] + b\left[{(\openone - F) \over 2}\right] + c\left[{(\openone + F) \over 2} - |+\rangle \langle +| \right], \nonumber
\end{eqnarray}
where $F= \sum |ij\rangle \langle ji|$ is the swap operator,
$a,b,c$ are real coefficients, and the three terms in square
brackets are mutually orthogonal projections. Hence a full
measurement obeying the $OO$ symmetry with $N$ outcomes is
characterised completely by the triples $(a_{k},b_{k},c_{k})$,
corresponding to each of the POVM elements indexed by $k=1,...,N$.
One could try to look for local protocols to match these triples
in a similar spirit to the solution presented for the Isotropic
case.  To be specific, it might be hoped that  a linear invertible
transformation could be found taking the $(a_{k},b_{k},c_{k})$ to
new triples $(x_{k},y_{k},z_{k})$, such that the
 $(x_{k},y_{k},z_{k})$ are the positive coefficients of some local protocol iff the
original $(a_{k},b_{k},c_{k})$ correspond to a PPT POVM.
However, we can show that no such naive transformation exists for the $OO$ symmetries (see
appendix 1 for details).
Consequently, in the following we will instead restrict our attention to the
{\it extreme points} of the convex set of $OO$ symmetric PPT POVMs. We will
present local protocols that attain these extreme points, and thereby argue that
all $OO$ symmetric PPT POVMS can be attained locally. First we need to
show how to construct these extreme points.

\subsubsection{Extreme points of the $OO$ symmetric POVMs}

\begin{figure}[hbt]
\begin{center}
    \leavevmode
    \epsfysize=6cm
    \epsfbox{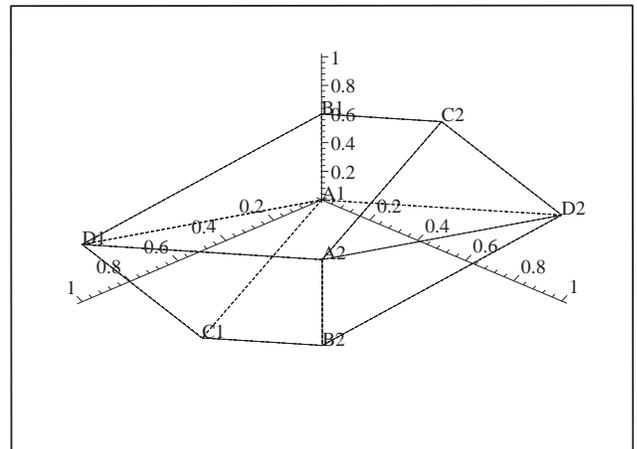}
    \caption{ The polyhedron of allowed PPT 2-outcome POVMs under $OO$ symmetry, sketched for $d=3$.
The vertices are given in equation (\ref{2outcome}).}
    \label{fig1}
\end{center}
\end{figure}

First let us discuss the partial transposition
of $OO$ symmetric operators. Partial transposition takes $M$ to an operator of the same form,
with new coefficients given by:
\begin{equation}
        \thvec{a'}{b'}{c'} = R \thvec{a}{b}{c}.
\label{opart}
\end{equation}
where
\begin{equation}
    R = {1 \over
    2d}\thmat{{2}}{{d(1-d)}}{{(d+2)(d-1)}}{{-2}}{{d}}{{(d+2)}}{{2}}{{d}}{{(d-2)}}.
\end{equation}
In order to satisfy the constraints of being a PPT POVM, all the
triples $(a_{k},b_{k},c_{k})$ corresponding to each element $M_k$
must satisfy:
\begin{eqnarray}
    a_k , b_k ,c_k \geq 0 \mbox{ (positivity)} \nonumber \\  a'_k ,b'_k ,c'_k \geq  0  \mbox{ (PPT-ness)} \nonumber \\ \sum_k a_k = \sum_k b_k = \sum_k c_k = 1 \mbox{ (normalisation)}.
\end{eqnarray}
We would like to construct a set of basic vectors for the $OO$
symmetries. One way to do this is to first construct the set of
2-outcome PPT extrema under these symmetries. In such situations
given one element $M$ the other is fixed as $\openone -M$ by
completeness. We hence require:
\begin{eqnarray}
   \openone\geq M \geq 0 \nonumber \\
   \openone\geq M^{\Gamma} \geq 0 .\nonumber
\end{eqnarray}
Denoting the only free element by a column vector of its
triple $M=(a,b,c)^{T}$, the constraints will result in a polyhedron
in $R^{3}$, for which the extreme points are the vertices (see figure \ref{fig1}).

It is trivial but tedious to calculate
these vertices. They are given
by the following POVM elements:
\begin{eqnarray}
    && A1 :\thvec{0}{0}{0} ~~~~~~ A2: \thvec{1}{1}{1} \nonumber \\
    && B1 : \thvec{0}{0}{2d/[d+2][d-1]} \nonumber \\
    && B2 : \thvec{1}{1}{(d+1)(d-2)/[d+2][d-1]} \nonumber \\
    && C1 : \thvec{1}{1/[d-1]}{(d-2)/[d+2][d-1]} \nonumber \\
    && C2 : \thvec{0}{(d-2)/[d-1]}{d^2 /[d+2][d-1]} \nonumber \\
    && D1 : \thvec{1}{0}{2/[d+2]} ~~~~~~ D2: \thvec{0}{1}{d/[d+2]} \label{2outcome}
\end{eqnarray}
Those POVM elements labelled with the same upper case letter are
{\it complementary}, in the sense that they sum to the identity. Hence each
letter labels a complete extremal PPT $OO$ symmetric POVM of 2
non-zero outcomes. The entire set of vectors in equation
(\ref{2outcome}) hence also
forms a set of basic vectors for $OO$ symmetries. We would like
to use this set to construct the entire set of extremal points. We
require any candidate extremal POVM to consist of measurement
elements that are linearly independent. Therefore we need only
restrict our attention to POVMs with at most 3 non-zero outcomes,
i.e. those measurements that can be represented by a $3 \times 3$
matrix:
\begin{equation}
    s = \thmat{{a_1}}{{a_2}}{{a_3}}{{b_1}}{{b_2}}{{b_3}}{{c_1}}{{c_2}}{{c_3}},
\end{equation}
where each column characterises one POVM element. As a consequence
of Lemma 7, we can only use three of our basic vectors to
construct the POVM, and as we require all three columns to be
non-zero, each column of the extremal 3-outcome POVMs must be {\it
proportional} to one and only one distinct basic vector. As there
are essentially only 7 important basic vectors to choose from in
the set above ($A1$ is trivial), we are already constrained as to the
possible form. However, there is another observation that will
provide much tighter constraints:

\smallskip

\noindent {\bf Lemma 17.} It is not possible to have 2 basic vectors
from the same complementary pair contributing non-zero weights in
the same extremal POVM. Consequently, no extremal POVM
can have any column with non-zero weight from the identity
basic vector, unless it is the trivial POVM with only
one non-zero outcome.

\smallskip

\noindent {\bf Proof.} Let us suppose that the two complementary basic
vectors are $V_1$ and $V_2$ respectively. Then
suppose that two POVM elements $j,k$ (where we can have $j=k$)
contain weights $p_1^j$ and $p_2^k$. Suppose that $p_1^j \leq p_2^k $.
Then we can write the whole POVM as a convex combination with
probability $p_1^j$ of a two-outcome POVM ($M_1 = V_1 , M_2 = V_2 $)
and $1 - p_1^j$ of the rest of the POVM rescaled to be complete. Hence
the POVM cannot be extremal unless one of the $p_1^j , p_2^k$ is zero $\square$.

\smallskip

An immediate consequence of Lemma 17 is that we cannot
use either of the basic vectors $A1, A2$ to construct our
POVM. Another consequence is that from the remaining
possible basic vectors, we must utilise exactly
one vector from each of the pairs $\{B1,B2\}$, $\{C1,C2\}$
and $\{D1 , D2\}$. We also require that in choosing
each basic vector, and its weight in a column, we must respect
completeness. Therefore it is not possible to choose all
three selected basic vectors with a 1 in the top
component. This leaves only six possible choices
for the basic vectors with non-zero weight:

\begin{eqnarray}
\{B1,C1,D1 \} \nonumber \\
\{B1,C1,D2 \} \nonumber \\
\{B1,C2,D1 \} \nonumber \\
\{B2,C1,D2 \} \nonumber \\
\{B2,C2,D1 \} \nonumber \\
\{B2,C2,D2 \}
\end{eqnarray}

It is not difficult, although tedious, to verify
that the only combination that can be made complete
with positive weights is the combination
$\{ B1 ,C1 ,D2 \}$. We recommend the use of a standard
computational package for algebraic manipulation,
with which it can readily be verified that
the selection $\{B2,C1,D2 \}$ is linearly
dependent, and that the remaining possibilities
other than $\{B1,C1,D2 \}$ all require negative
weights in order to satisfy completeness \cite{request}. For
$\{ B1 ,C1 ,D2 \}$  the weights are also fixed uniquely,
as the solution involves a set of 3 linearly
independent equations in 3 unknowns. It turns
out that two of the three weights are
1 (those of $B1$ and $C1$), and as these
elements are also 2-outcome extrema, they
cannot be perturbed in any way while keeping
the whole POVM feasible. Hence this whole 3-outcome
POVM is extremal, and we have the following:

\smallskip

\noindent {\bf Lemma 18.} The only genuine 3-outcome extremal POVM for $OO$
symmetries, upto
relabelling the outcomes, is given by the POVM with elements
represented by the following triples:
\begin{eqnarray}
M_1 &=& (0,0,{ 2d  \over (d+2)(d-1)}) \nonumber \\
M_2 &=& (0,{d-2\over d-1},{ d(d-2)  \over (d+2)(d-1)  }) \nonumber \\
M_3 &=& (1,{1 \over d-1},{ d-2  \over (d+2)(d-1)}) , \label{3outcome}
\end{eqnarray}
where we have made the
slight abuse of notation that a POVM element will be set equal
to the triple representing it.

\subsubsection{Locally attaining the extreme points of the $OO$ symmetric
POVMs}

In trying to attain the extreme points derived above,
we will first consider only the 2-outcome extrema, given
in equation (\ref{2outcome}).The important question is how to realise the extreme points of this polyhedron
using LOCC measurements. The measurement $A$ and its complement are trivial,
they correspond to the do nothing POVM. The
measurement $D$ and its complement are also not too difficult to obtain.
It is easy to confirm that if Alice and Bob carry out an `orthogonal twirl' \cite{isotwirl,lookup} followed by the
locally implementable projection:
\begin{equation}
\sum_{i=0}^{d-1}|ii \rangle \langle ii|
\end{equation}
then they will achieve the POVM $D$, and so both $D$ and its complement can be attained.

The points $B,C,$ and complements are a little more subtle. However,
by performing an orthogonal twirl \cite{lookup} it is not difficult to show
 that if we can find a set of overcomplete pure states on Alice's
particle $\{|q\rangle \}, q=1...L, L \geq d$ with the properties
that:
\begin{eqnarray}
{d \over L}\sum_{q=1}^{L} |q\rangle \langle q| &=& I_{A} \label{qcom}\\
\forall~q ~~~~ \mbox{tr} \{|q \rangle \langle q| (|q \rangle \langle q|)^{T}    \} &=& 0 \label{qorth}
\end{eqnarray}
then the following local measurements:
\begin{equation}
{d \over L}\sum_{q=1}^{L} |q\rangle \langle q| \otimes |q\rangle \langle q|  \nonumber
\end{equation}
and
\begin{equation}
{d \over L}\sum_{q=1}^{L} |q\rangle \langle q| \otimes (|q\rangle \langle q|)^{T}  \nonumber
\end{equation}
will attain points $B$ and $C$ respectively when preceded by the orthogonal twirl.
It remains to show that a set of pure states satisfying equations (\ref{qcom}) and (\ref{qorth})
exists. Two pure states with property (\ref{qorth}) are:
\begin{eqnarray}
 |u\rangle := 1/\sqrt{2} (  |r\rangle + i |s\rangle) \nonumber \\
 |v\rangle := 1/\sqrt{2} (  |r\rangle - i |s\rangle)
\end{eqnarray}
for any two computational basis states $ |r\rangle , |s\rangle $. Suppose
that $d$ is even. Then Alice can obtain a complete set of  $\{|q\rangle \}$
satisfying the properties by partitioning her space in 2-level orthogonal subspaces,
and constructing a $ |u\rangle ,|v\rangle $ for each subspace. If $d$ is odd,
then she can do the same until she is left with a 3-level subspace. We therefore
need to construct such a set of pure states for this 3-level subspace. One route
to a solution is as follows. Without loss of generality
assume that this residual subspace is spanned by computational basis states $ |1\rangle , |2\rangle , |3\rangle $.
Then Alice can pick any irreducible representation of a finite group in
$3 \times 3$ real orthogonal matrices, i.e. $OO^{T}=I$. Suppose this set is $\{O_{m}|m=1...L\}$.
Then the set of pure states:
\begin{equation}
O_{m} [ 1/\sqrt{2} (  |1\rangle + i |2\rangle)]
\end{equation}
will satisfy the requirements (\ref{qcom}) and (\ref{qorth}). It seems likely that
this solution for odd $d$ is overly complicated, however, we were not able to find any
significant simplifications.

Therefore we see that all extreme points of the set of 2-outcome PPT $OO$ symmetric measurements can
be obtained by local protocols involving only one way communication. Now we need to
turn to the 3-outcome element presented in equation (\ref{3outcome}). Given the
existence of a set of pure states $\{|q\rangle | q=1...L, L \geq d \}$ satisfying
the conditions (\ref{qcom}) and (\ref{qorth}), it is not too difficult to verify
that the local POVM deifened by the following elements:
\begin{eqnarray}
 N_1 &=&  {d \over L}\sum_{q=1}^{L} |q\rangle \langle q| \otimes |q\rangle \langle q| \nonumber \\
 N_2 &=&  {d \over L}\sum_{q=1}^{L} |q\rangle \langle q| \otimes (|q\rangle \langle q|)^{T} \nonumber \\
 N_3 &=&  {d \over L}\sum_{q=1}^{L} |q\rangle \langle q| \otimes [\openone _B - |q\rangle \langle q| - (|q\rangle \langle q|)^{T}]
\end{eqnarray}
achieves the $M_1 , M_2 , M_3 $ of eq.(\ref{3outcome}) respectively when preceded by the orthogonal twirl.

Therefore all extreme points for N-outcome OO symmetric POVMs can be attained
by local protocols.

\section{Summary $\&$ Discussion}

The classification of Separable and PPT operations seems to be
particularly suited to the question of measurements, where
we find that a POVM can be implemented by a separable operation
iff its elements are separable, and by a PPT operation
iff its elements are PPT. As the Separable/PPT POVMs are mathematically
more tractable than the LOCC measurements, and as we have the (strict) inclusions
LOCC $\subset$ Separable $\subset$ PPT $\subset$ Global,
these classes of operation are useful for deriving bounds on what we can achieve
with LOCC observations. For this reason it is useful to have techniques for
deriving extremal measurements under the constraints of separability
and PPT-ness. In this context we have discussed a number of
useful tools. The partial transposition isomorphism is particularly
useful for PPT measurements under partially conjugated
symmetries, whereas the linear independence requirements and
basic vectors approach are more generally useful for extremising
POVMs under any homogeneous constraints. Applying
these techniques we were able to show that for Isotropic, Werner, Bell and $OO$
symmetries, the separability of POVM elements is necessary
and sufficient for local implementability. There is no
obvious reason why this is the case, as it is clear from
the results of \cite{nonlocalitywithoutentanglement} that not
all separable POVMs can be implemented locally.

Perhaps the most immediate consequences of our results are for
local state discrimination. Imagine that we have been
given one of $i$ states $\rho_i$ with probability $p_i$.
What is the best way to tell which state we have been
given? Questions such as this are significant for
entanglement distillation protocols, where `target' states are
often locally measured to give information about `source' states \cite{Benndist,Volldist},
and also for cryptographic
schemes. In such scenarios one expects the extremal local
POVMs to be the optimal possible measurements for most
reasonable cost functions, regardless of the number of states
or prior probabilities. Therefore our results also essentially
give optimal protocols for the local discrimination
of Isotropic, Werner, Bell and $OO$ states. In the context of state
discrimination the Bell symmetries give the largest (in general)
separation between local and global state discrimination. If
we have global access to one of the four orthogonal
Bell pure states, then we can discriminate them
perfectly, potentially obtaining 2 classical bits
of information. However, if we are restricted to acting
locally, the best measurement we can do has at most
2 non-zero outcomes, showing that we can locally
obtain at most 1 bit of information. Unfortunately
this does not supply an interesting bit hiding scheme
in the manner of \cite{hb1,eggeling}, as the technology required
to implement the scheme is the same as the technology
required to break it - Alice and Bob can share prior Bell states
and use them to teleport Alice's state to Bob, thereby
allowing Bob full access to the state without having to meet Alice. Nevertheless,
our results give the first examples of mixed states
for which a full quantitative analysis of optimal
state discrimination can be performed.

We hope to be able to apply the techniques from this paper
to more ambitious symmetries, perhaps considering more
specific cost functions or prior probabilities in order
to make the calculations more tractable. A fuller understanding
of local measurements could have applications not
only for local state discrimination and entanglement
distillation, but also for any situation
in which local observations are required to demonstrate
a phenomenon, such as in demonstrations of non-locality,
and also the recent program of research on the experimental
implementation of entanglement witnesses \cite{guhne}.

\section{Acknowledgements}

SV thanks Oliver Rudolph, Adam Brazier and Tilo Eggeling for sharing their insights,
and Paolo Lo Presti $\&$ Mauro D'Ariano for discussions about
global extremal POVMs. We acknowledge financial support from EC project EQUIP (IST-1999-11053),
EC project ATESIT (contract IST-2000-29681), EC project ATESIT
(contract IST-2000-29681), INFM PRA 2001 CLON, US Army Grant DAAD 19-02-0161,
the UK EPSRC and the European Science Foundation Programme on
Quantum Information Theory and Quantum Computing.

\section{Appendix 1: No `naive' solution possible for $OO$ symmetries.}

In this appendix we will see that no `naive' solution can be
found for $OO$ symmetric POVMs with the same simplicity as the
solution for the Isotropic/Werner measurements. First we need
to clarify what we mean by `naive'. We will consider local
measurement protocols consisting of POVM elements $N_k$,
each of which can be written in the following form:
\begin{equation}
N_k = x_k X + y_k Y + z_k Z
\label{stip}
\end{equation}
where the $\{X,Y,Z\}$ are orthogonal projectors that sum to the identity,
and the ($x_k$, $y_k$, $z_k$) are triples of real coefficients.
Then the requirements of completeness and positivity mean that
the sets of $\{x_k \}$, $\{y_k \}$ and $\{z_k \}$ must form probability distributions.
We will also assert that {\it any such valid POVM is also local}. Under
these assertions we will also impose the following requirement:

\begin{itemize}
\item Any PPT POVM of $OO$ symmetry can be attained by performing
an orthogonal twirl followed by a (local) measurement of the
stipulated form (\ref{stip}).
\end{itemize}
We will show that no such solution
is possible for the $OO$ case, even though a solution
with essentially the same features was possible for the Isotropic
case.

Let us represent each POVM element $N_k$ by the column vector
$\vec{v_k} = (x_k , y_k , z_k )^T$. Then under twirling, $N_k$ will be
taken to an $OO$ symmetric POVM element. Let this element
be represented by the column vector $\vec{w_k} = (a_k , b_k , c_k )^T$,
using the same notation as the main text. There will be
a linear transformation $L$, such that $L \cdot \vec{v_k} = \vec{w_k}$
represents the effect of the twirling.

As stated in the main text, the the only non-zero elements
in extremal (PPT) POVMs must be linearly independent (Theorem 5). Let us
represent the set of general extremal POVMs of the stipulated form (\ref{stip}) by
the set $\{r_i\}$, and the extrema of the PPT $OO$ symmetric form by the set $\{s_i\}$.
As a consequence of the linear independence requirement,
every one of these extrema can be expressed as $3 \times 3$
matrices, where the columns are linearly independent and represent
each of three non-zero POVM elements. Hence two general members
$r \in\{r_i \}$ and $s \in\{s_i \}$ of these sets can be written:
\begin{equation}
r = \thmat{{x_1}}{{x_2}}{{x_3}}{{y_1}}{{y_2}}{{y_3}}{{z_1}}{{z_2}}{{z_3}},
~s = \thmat{{a_1}}{{a_2}}{{a_3}}{{b_1}}{{b_2}}{{b_3}}{{c_1}}{{c_2}}{{c_3}}.
\end{equation}
As the columns of these matrices are linearly independent, the matrices
are invertible. This will be important in the subsequent discussion.

Now let us see the implications of the above assertions
for the form of the transformation $L$. We are forced to have the following properties:
\begin{itemize}
\item $L$ must have non-negative matrix elements, as its elements are calculated
from the traces of products of non-negative operators.
\item Each row of $L$ must sum to 1. This is from the fact that
twirling retains completeness of the POVM.
\item $L$ must be invertible. As we wish {\it all} PPT
$OO$ POVMs to be obtainable from the stipulated protocol, the extreme points $\{s_i\}$
of the PPT $OO$ symmetric POVMs  must be contained within
set obtained from $L$ acting on the extrema $\{r_i\}$ of the stipulated
measurement. Consequently for some $r,s$ we have $L \cdot r = s$,
and hence $L^{-1}=r \cdot s^{-1}$.
\end{itemize}
Let us now consider the transformation $R$ of equation (\ref{opart}).
The vector:
\begin{equation}
R \cdot L \cdot\vec{v_k} = R \cdot \vec{w_k}
\end{equation}
gives the partial transpose of the twirled POVM element corresponding
to $N_k$. As we wish {\it any} PPT POVM be attainable by a
local protocol of our postulated form, and as we require
that $L$ be invertible, we also require that both:
\begin{equation}
L \cdot\vec{v_k} \geq ~ 0 ~ \forall ~ k
\end{equation}
and
\begin{equation}
R \cdot L \cdot\vec{v_k} \geq ~ 0 ~ \forall ~ k
\end{equation}
hold if and only if the sets $\{x_k \}$, $\{y_k \}$ and $\{z_k \}$
form probability distributions. We also must require the following
properties for the matrix product $R \cdot L$:

\begin{itemize}
\item  $R \cdot L$ must contain only non-negative elements. This can be seen as follows.
If we  pick a three element local measurement characterised by:
\begin{equation}
\vec{v_1} = (1,0,0)^{T} ~;~\vec{v_2} = (0,1,0)^{T} ~;~\vec{v_3} = (0,0,1)^{T}
\end{equation}
then this must produce vectors $\vec{w_k}$ with non-negative partial transposition, and
it can readily be verified that this implies that the elements of $R \cdot L$
must be non-negative.
\item The rows of $R \cdot L$ must sum to 1. This can be shown from the fact that
the same property holds for both $R$ and $L$.
\end{itemize}
Now consider the following disallowed choice for the vectors
$\vec{v_k}$, where $\epsilon$ is small and positive:
\begin{eqnarray}
\vec{v_1} = (1/3,1/3,-\epsilon)^{T} &~;~& \vec{v_2} = (1/3,1/3,(1 + \epsilon)/2 )^{T} ~; \nonumber \\
\vec{v_3} = (1/3,&1/3&,(1 + \epsilon)/2 )^{T}.
\end{eqnarray}
This choice of vectors does not give a valid choice of measurement vectors, as the first
one contains a negative component. However, the vectors $\vec{v_2}$ and $\vec{v_3}$ can in principle
come from valid probability distributions by themselves, and so it must be the case
that either $L \cdot \vec{v_1}$ or $R \cdot L \cdot \vec{v_1}$ contains a negative component. It can be shown
that this, together with the fact that both $L$ and $R \cdot L$ must be
row stochastic, implies that at least one row of either $L$ or $R \cdot L$ must be $(0,0,1)$. Similarly
we can `place the $\epsilon$' in different rows in the above vectors to show
that in fact we require that at least one row of  either $L$ or $R \cdot L$ must
be $(0,1,0)$ and at least one row should be $(1,0,0)$. Regardless
of where these rows appear, as $R=R^{-1}$ contains some negative matrix elements, no
such solution is possible, as both $L$ and $R \cdot L$ are required to have
non-negative matrix elements. Although the proof that we have given is tailored
to $OO$ symmetries, it can be modified relatively easily for other situations,
including bases of Bell diagonal states.

\end{multicols}

\begin{references}
\bibitem{commutant}
The commutant of a group $g$ of unitaries is defined as the set of
all operators $A$ for which $[U,A]=0$ for all $U\in g $.
\end{references}
\end{document}